\documentclass[preprint,amsmath,amssymb,prl,lengthcheck]{revtex4-2}
\usepackage{graphicx,color,hyperref,verbatim}
\usepackage{dcolumn}
\usepackage[normalem]{ulem}

\usepackage{bm}

\begin{document}

\title{First-Principles Theory of the Relativistic Magnetic Reconnection Rate in Astrophysical Pair Plasmas}

\author{Matthew Goodbred}
\affiliation{Dartmouth College \\
Hanover, New Hampshire 03755, USA}

\author{Yi-Hsin Liu}
\affiliation{Dartmouth College \\
Hanover, New Hampshire 03755, USA}

\begin{abstract}
We develop a first-principles model for the relativistic magnetic reconnection rate in strongly magnetized pair plasmas. By considering the energy budget and required current density near the x-line, we analytically show that in the magnetically-dominated relativistic regime, the x-line thermal pressure is significantly lower than the upstream magnetic pressure due to the extreme energy needed to sustain the current density, consistent with kinetic simulations. This causes the upstream magnetic field lines to collapse in, producing the open outflow geometry which enables fast reconnection. The result is important for understanding a wide range of extreme astrophysical environments, where fast reconnection has been evoked to explain observations such as transient flares and nonthermal particle signatures. 

\end{abstract}
\keywords{Magnetic reconnection}

\maketitle

\begin{figure}[ht!]
    \centering
    \includegraphics[width=3 in]{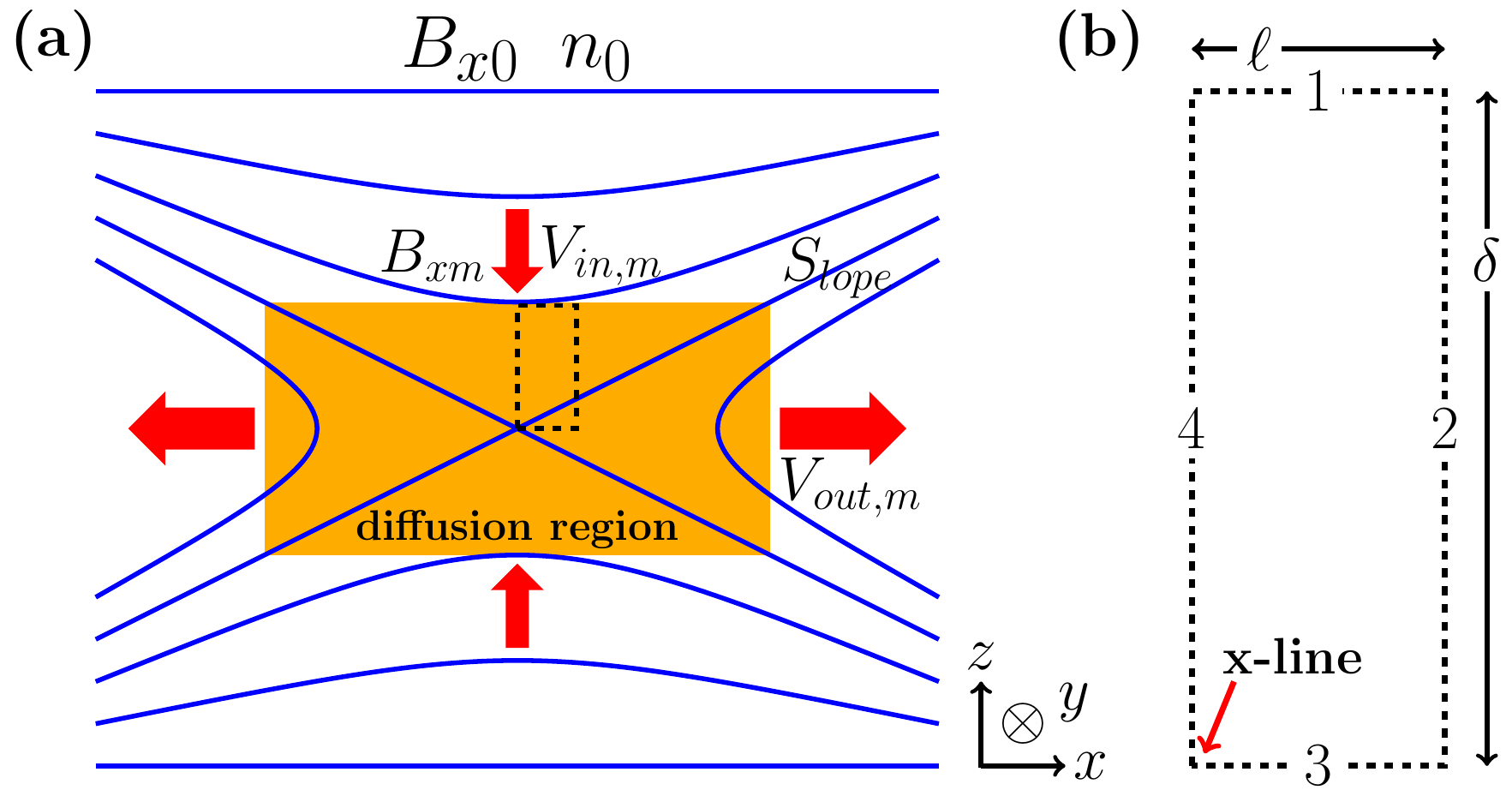}
    \caption{Panel (a) depicts the reconnection geometry, including the distinction between asymptotic and microscale quantities. Panel (b) is a blow-up of the dotted box in panel (a), the Gaussian surface to the upper right of the x-line. }
    \label{fig:cartoon}
\end{figure}

{\it Introduction--} Magnetic reconnection is a fundamental plasma phenomenon which explosively converts magnetic energy into plasma energy.
Inflowing plasma carries magnetic flux into the {\it diffusion region}. 
There plasma is energized and ejected around the {\it x-line} into the exhaust (see Fig.~\ref{fig:cartoon}). Reconnection is thought to play a critical role in ubiquitous astrophysical environments where the magnetic field energy density can exceed the rest mass energy density of the ambient plasma \citep{Kagan2015}, characterized by the asymptotic magnetization parameter $\sigma_0\equiv B_{x0}^2/4\pi n_0 mc^2$. Here $B_{x0}$ and $n_0$ are the asymptotic (background) reconnection magnetic field strength and particle density, respectively. When $\sigma_0\gg 1$, relativistic effects become important. The reconnection rate is the most important characteristic of magnetic reconnection since it describes how fast reconnection processes magnetic flux and indicates the particle acceleration timescale. 
Fast reconnection, on the order of $R_0\sim0.1$ in normalized units, has been evoked to explain transient gamma-ray flares in active galactic nuclei (AGNs) \citep{Sironi2015,Ripperda_2022}, the dissipation of magnetic `hair' in black hole (BH) magnetospheres \citep{Bransgrove2021}, BH accretion disk coronal heating and flares \citep{DiMatteo1998}, gamma-ray bursts (GRBs) in pulsar nebula \citep{Cerutti2014,Lyutikov2018}, the rapid dissipation of Poynting flux in pulsar winds \citep{Porth2013,Cerutti2020}, and giant magnetar flares \citep{Masada2010}. 
Electron-positron (pair) plasmas are thought to dominate the plasma environment of pulsar winds \citep{Cerutti2012} and play a significant role in BH magnetospheres, including AGN/BH jets \citep{Wardle1998,Crinquand2020}. Highly magnetized electron-ion (e-i) plasmas may be relevant in BH accretion discs \citep{DiMatteo1998} and TeV flares in relativistic AGN jets \citep{Zabandan2012, Nalewajko2011,Sironi2015}. In these contexts, reconnection can accelerate particles to ultra-relativistic energies and produce broad nonthermal spectra \citep{Guo2020,Guo2014,Sironi2014,Uzdensky2022}. 
Despite the great interest in pair and relativistic e-i reconnection to explain numerous puzzling observations in astrophysics, there is yet no first-principles theory for the rate of reconnection in these astrophysical plasmas \citep{Kagan2015}. 

It was recently shown that in nonrelativistic ($\sigma_0\ll 1$) e-i plasmas, reconnection is fast because Hall electromagnetic fields divert inflowing Poynting flux around the x-line. Electrons remain frozen-in within the ion diffusion region, and as the primary out-of-plane current carrier, they drag reconnecting magnetic field lines into the out-of-plane Hall quadrupole magnetic field. Inflowing electrons are subsequently deflected by this Hall magnetic field into the outflow direction, carrying the magnetic energy away from the x-line.
The resulting energy void prevents thermal pressure build-up at the x-line, and upstream magnetic field lines bend in to maintain force balance along the inflow \citep{Liu2022}. This creates the open outflow geometry necessary for fast reconnection \citep{Liu2017}. The x-line pressure depletion is also observed during fast reconnection in relativistic pair plasmas \citep{Liu2020}. However, the cause of this depletion must be drastically different since the plasma species' equal mass eliminates the Hall effect, and all inflowing magnetic energy can be transferred to the plasma at the x-line. Similarly, in the large $\sigma$ limit of e-i reconnection, the scale separation between species may be less relevant since the effective relativistic masses of electrons and ions become equivalent, indicating that relativistic e-i reconnection behaves similarly to pair plasma reconnection where the Hall effect is absent \citep{Werner2018}. 

In this Letter, we identify an entirely different mechanism leading to fast reconnection in relativistic pair plasmas. We show that in the magnetically-dominated relativistic regime, the x-line thermal pressure cannot balance the upstream magnetic pressure due to the energy needed to sustain the extreme current density. The implosion of the upstream magnetic pressure into the x-line triggers fast reconnection. With this insight, we develop a first-principles model for the reconnection rate in strongly magnetized pair plasmas with a simple magnetic field reversal.
The model predictions compare well to fully kinetic particle-in-cell (PIC) simulations.

\begin{figure*}[thb!]
    \centering
    \includegraphics[width=6.4 in]{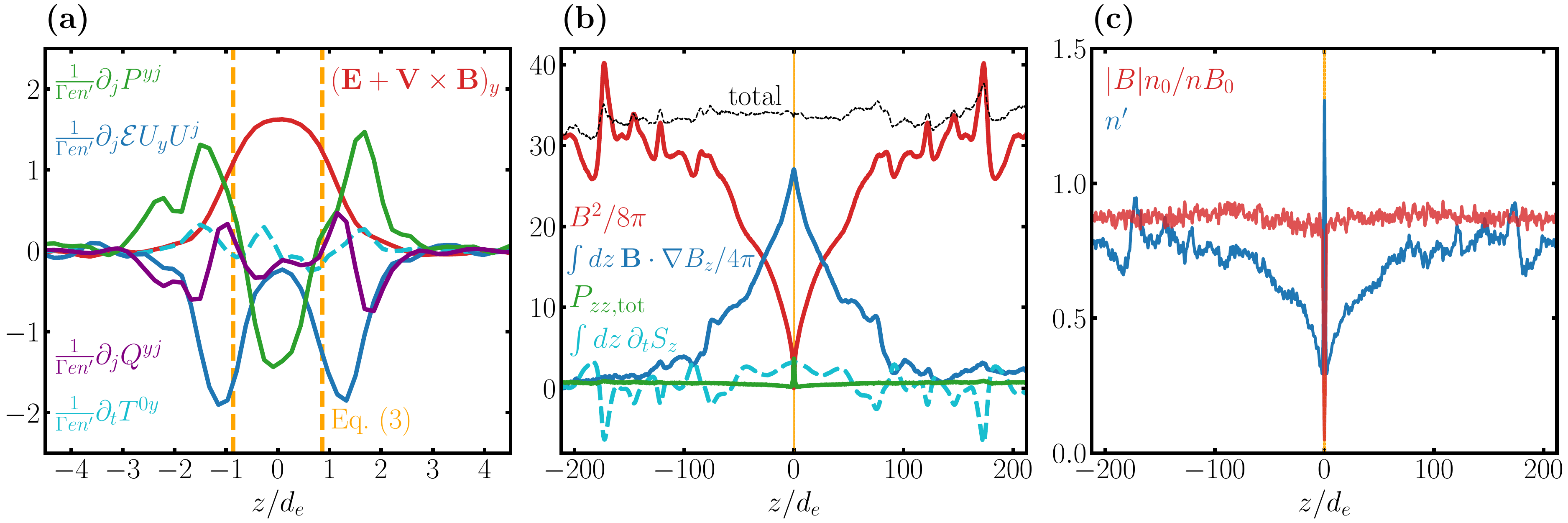}
    \caption{Panel (a) shows the out-of-plane electron Ohm's law around the x-line. Panel (b) plots the total pressure balance including both species along the inflow symmetry line, demonstrating that $P_{zz}|_{xline}\ll B_{x0}^2/8\pi$. The Poynting flux time derivative balances fluctuations far upstream.  Panel (c) shows the electron proper density $n^\prime$ and the normalized $|B|/n$ ratio along the inflow symmetry line. Orange vertical bars show the prediction of the diffusion region thickness based on Eq.~(\ref{eq:delta}). All plots are from the $\sigma_{0,\text{init}} =100$ simulation run at $t=379/\omega_{p}$, but similar features are observed in all runs.}
    \label{fig:ohms}
\end{figure*}
{\it Theory--} Our model of the x-line plasma pressure considers two basic properties of pair plasma reconnection: the energy budget around the x-line and the current density necessary for the field reversal. As in \citet{Liu2022}, these properties are analysed within a Gaussian surface with the bottom-left corner at the x-line, illustrated by the dotted rectangle in Fig.~\ref{fig:cartoon}. The surface extends to the current sheet half-thickness $\delta$ in the z-direction and has small width $\ell\ll\delta$ in the x-direction, as shown in Fig.~\ref{fig:cartoon}(b). 
The inflow, outflow, and out-of-plane directions are z, x, and y, respectively. The asymptotic background conditions of magnetic reconnection can be very different than the microscale conditions at the edge of the diffusion region, as illustrated in Fig.~\ref{fig:cartoon}(a) \citep{Liu2017}. Thus asymptotic and microscale quantities are denoted with `$0$' and `$m$' subscripts, respectively. We will also refer to local quantities around the diffusion region by the labelled surfaces in Fig.~\ref{fig:cartoon}(b). Note that surface 1 is at the microscale. Electron and positron species are specified by $(-)$ and $(+)$ subscripts, but due to the symmetry between species, we omit subscripts unless relevant.

Quantities in the proper frame of a species are primed.
Roman letters index the 3-dimensional Euclidean space ($i=x,y,z$). Greek letters index the 4-dimensional flat space-time ($\alpha=0,x,y,z$), and we use the mostly-positive metric tensor $\eta^{\alpha\beta}=\text{diag}(-1,1,1,1)$. The speed of light $c=1$, but we write the unit where instructive. In the following analysis, we follow the tensor formalism of \citet{Zenitani2018}. 
The stress-energy (SE) tensor is obtained from the particle 4-velocity $u^\alpha=(\gamma, \gamma {\bf v})$ and Lorentz factor $\gamma=1/[1-v^2]^{1/2}$ as 
\begin{equation}\label{eq:se_def}
    T^{\alpha\beta}=m\int f({\bf u})u^\alpha u^\beta \frac{d^3 u}{\gamma},
\end{equation}
which is a 4-tensor because it is a linear combination of the dyadic 4-tensor $u^\alpha u^\beta$ and since the distribution function $f(\bf{u})$ and $d^3u/\gamma$ are both Lorentz invariant. The normalization of $f(\bf{u})$ is defined by $\int{f(u)d^3u=n}$. Although generally not explicitly denoted, $f(\bf{u})$ is also a function of location. 

For the remainder of this work, all velocities refer to the Eckart velocity. The Eckart 4-velocity $U^\alpha_E\equiv N^\alpha/n^\prime$ where $N^\alpha \equiv \int f({\bf u}) u^\alpha \frac{d^3 u}{\gamma}$ is the particle number flux 4-vector and $n^\prime=(-N^\alpha N_\alpha)^{1/2}$ is the proper density \citep{Eckart1940}. In the Eckart frame, the spatial components $U_E^i=0$, and thus the spatial components of the particle number flux 4-vector vanish (i.e., $N^i=0$), indicating the rest frame of the bulk fluid. Similarly, the Eckart 3-velocity $V^i_E\equiv U^i_E/\Gamma_E$ with $\Gamma_E\equiv 1/[1-V_E^2]^{1/2}$ the Lorentz factor of the Eckart frame. Hereafter, we set $U^\alpha=U_E^\alpha$ and $V^i = V^i_E$.

With this background, the relativistic electron Ohm's law is written as 
\begin{equation}\label{eq:Ohms_Law}
    \mathbf{E}+\mathbf{V}\times\mathbf{B}=-\frac{1}{\Gamma e n^\prime}\left[\partial_j\left(\mathcal{E}U^i U^j+Q^{ij}+P^{ij} \right)+ \partial_t T^{i0} \right]
\end{equation}
where $Q^{ij}$ is the heat flux tensor and $P^{ij}$ is the pressure tensor \citep{Zenitani2018}. See the Appendix for the full derivation. 
It is known that in the relativistic regime, the bulk inertia term of Ohm's law balances the reconnection electric field $E_y$ at the edge of the diffusion region \citep{Hesse2007,Zenitani2018,Xiong}, as seen by comparing the blue to red curve in Fig.~\ref{fig:ohms}(a).
In the Appendix, we evaluate Eq.~(\ref{eq:Ohms_Law}) at the transition region close to surface 1 and find that the current sheet width is approximately the pair plasma inertial length based on the proper density inside the current sheet:
\begin{equation}\label{eq:delta}
    \delta \approx \sqrt{\frac{mc^2}{8\pi n_{3}^\prime e^2}}.
\end{equation}
This prediction is validated in Fig.~\ref{fig:ohms}(a) and has been noted in previous works \citep{ Treumann2011,Cerutti2012}. Close to the x-line, the relativistic compression of the plasma in the current sheet is based on the Lorentz factor of the bulk flow in the y-direction, $\Gamma_y \equiv 1/[1-V_{y3}^2]^{1/2}$, so that Ampere's law is $B_{xm}/4\pi\delta =2en_3^\prime\Gamma_yV_{y3}$. Plugging in Eq.~(\ref{eq:delta}), we solve for
\begin{equation}\label{eq:gammay}
    \Gamma_{y} \approx \sqrt{1+\frac{\sigma_m}{2}\left(\frac{n_{1}}{n_{3}^\prime}\right)}
\end{equation}
where $\sigma_m=B_{xm}^2/4\pi n_{1}mc^2$ is the microscale magnetization parameter.

Next, we consider 
the energy available to support the x-line plasma pressure.
The energy conservation equation is obtained from the vanishing 4-divergence of the time component of the total SE tensor:
\begin{equation}\label{eq:energy_conserv}
    \partial_\alpha \left(T_{(+)}^{0\alpha}+T_{(-)}^{0\alpha }+T_{EM}^{0\alpha}\right)=0
\end{equation}
where $T_{EM}^{\alpha\beta}=(1/4\pi)\left[F^{\alpha\mu}F^{\beta}_{\;\;\mu}-\eta^{\alpha\beta}F_{\mu\nu}F^{\mu\nu}/4 \right]$ is the electromagnetic SE tensor. We wish to analyse this equation within the Gaussian surface indicated by the dotted box Fig.~\ref{fig:cartoon}. By symmetry around the x-line, the energy fluxes through surfaces 3 and 4 vanish. In steady state $\partial_t=0$, so we use the divergence theorem to rewrite Eq.~(\ref{eq:energy_conserv}) as 
\begin{align}\label{eq:energy_integral}
    & \int_{1}{\left(T_{(+)}^{0z}+T_{(-)}^{0z}+T_{EM}^{0z}\right)}dx\nonumber\\ + & \int_{2} \left(T_{(+)}^{0x}+T_{(-)}^{0x}+T_{EM}^{0x}\right)dz =0.
\end{align}
With a cold upstream plasma, $\gamma\simeq 1$ in the inflow region. Thus by Eq.~(\ref{eq:se_def}), 
\begin{equation}\label{eq:T_in}
    \left.\left(T_{(+)}^{0z}+T_{(-)}^{0z}\right)\right|_{1}\approx 2n_{1}mc^2 V_{z1}. 
\end{equation}

In this work we aim to analytically show that $P_{zz}|_{xline}$ (i.e, a thermal spread in $v_z$) is significantly lower than the asymptotic magnetic pressure in the large-$\sigma_0$ limit (as in the comparison of the green and red curves in Fig.~\ref{fig:ohms}(b)) because the current carrier bulk flow kinetic energy takes most of the incoming magnetic energy. While there is in reality some thermal spread, for the purposes of this model we assume the plasma is cold in the $v_y$ and $v_x$ directions, which minimizes the energy required to sustain the current. 
Since $d^3u=\gamma^5 d^3v$, we consider the distribution function $g\equiv\gamma^5 f=F(v_z)\delta(v_x-V_x)\delta(v_y-V_y)$, so that $g$ is the number density in 3-velocity phase space. Then
\begin{align}\label{eq:T_out}
\left.\left(T_{(+)}^{0x}+T_{(-)}^{0x}\right)\right|_{2} &\approx 2m\int{f_2 u_x c d^3u} =2m\int{g_2 \gamma v_x c d^3v} \nonumber \\
&= 2mV_{x2} c\int {F_2(v_z) \gamma(v_z)dv_z}  \nonumber \\
&=2\langle \gamma(v_z) \rangle_{2}n_{2}mc^2 V_{x2}.    
\end{align}
where we define $\gamma(v_z)\equiv \gamma(v_x=V_x, v_y=V_y, v_z)$ and the ensemble average $\langle A\rangle \equiv \int g A d^3v/\int g d^3v$.
Lastly, the inflowing electromagnetic energy flux $\left.T_{EM}^{0z}\right|_1=-E_{y}B_{xm}/4\pi$. The plasma is frozen-in in the upstream, so $E_y=-V_{z1}B_{xm}$. The result is
\begin{equation}\label{eq:TEM_in}
    \left.T_{EM}^{0z}\right|_1 =\frac{B_{xm}^2}{4\pi}V_{z1}.
\end{equation}
In the $|\partial_xB_z|\ll|\partial_zB_x|$ limit, we can igore the outflowing EM energy flux: $\left.T_{EM}^{0x}\right|_2\approx 0$. Combining Eqs.~(\ref{eq:T_in},~\ref{eq:T_out},~\ref{eq:TEM_in}) with Eq.~(\ref{eq:energy_integral}), we obtain
\begin{equation} \label{eq:energy}
    \left[2n_{1} mc^2 +\frac{B_{xm}^2}{4\pi} \right]V_{z1}\ell + 2\langle \gamma(v_z) \rangle_{2}n_{2} mc^2 V_{x2}\delta \approx 0.
\end{equation}
With the particle number continuity equation $n_{1}V_{z1}\ell + n_{2}V_{x2}\delta=0$, we solve Eq.~(\ref{eq:energy}) for
\begin{equation}\label{eq:gamma}
\langle \gamma(v_z) \rangle_{2} \approx 1+ \frac{\sigma_m}{2}.
\end{equation}

Finally, we can estimate the x-line pressure. At the mid-plane $V_z=0$, so the total $P_{zz}$ at surface 3 is given by 
\begin{align}
&\left.\left(T^{zz}_{(+)}+T^{zz}_{(-)}\right)\right|_3=2m\int f_3 \frac{u_z^2}{\gamma}d^3u=2m\int g_3 \gamma v_z^2d^3v\nonumber \\
&=2mc^2\int F_3\gamma(v_z)\left[1-V_{x3}^2-V_{y3}^2-1/\gamma(v_z)^2\right] dv_z \nonumber\\
&=2n_{3}mc^2\left[\frac{\langle \gamma(v_z) \rangle_{3}}{\Gamma_y^2}-\left\langle\frac{1}{\gamma(v_z)}\right\rangle_{3} \right]
\end{align}
where we used the fact that $V_{x3}\rightarrow 0$ in the $\ell \rightarrow 0$ limit. Using the inequality $\langle 1/A \rangle \geq 1/\langle A \rangle$ for $A>0$ and $n_{3}=\Gamma_yn_{3}^\prime$, we then arrive at
\begin{equation}\label{eq:P2}
 \left.P_{zz}\right|_{xline} \leq 2n'_{3}mc^2 \left[\frac{\langle \gamma(v_z)\rangle_{3}}{\Gamma_y}-\frac{\Gamma_y}{\langle \gamma(v_z)\rangle_{3}} \right].
\end{equation}

\begin{figure*}[htb!]
    \centering
    \includegraphics[width=6 in]{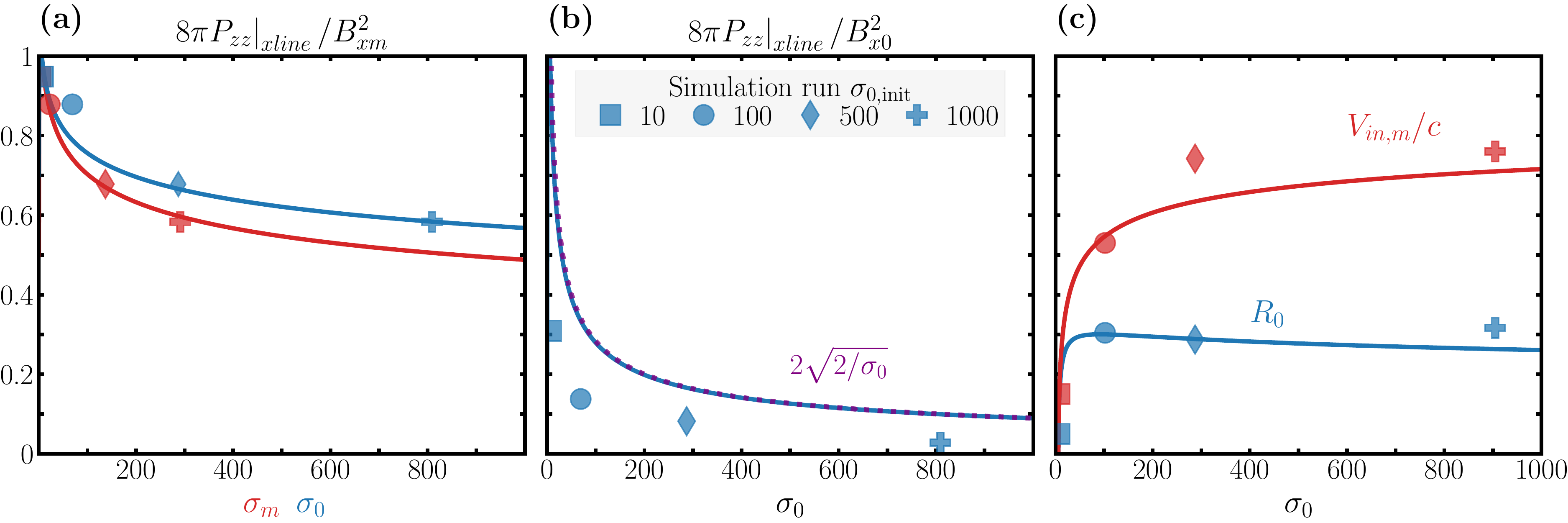}
    \caption{Panel (a) shows the microscale x-line pressure ratio as a function of both $\sigma_m$ and $\sigma_0$. Panel (b) shows the pressure ratio relative to the asymptotic magnetic pressure as a function of $\sigma_0$. The dashed purple curve is the $\sigma_0\gg 1 $ limit of the numerical solution, $2\sqrt{2/\sigma_0}$. Panel (c) shows the reconnection rate $R_0$ and the microscopic inflow speed $V_{in,m}/c\simeq E_y/B_{xm}$ as a function of $\sigma_0$. To measure $R_0$ from simulations, we use $E_y=\partial_t\left[\text{max}(A_y)-\text{min}(A_y) \right]$ where $A_y$ is the y-component of the vector potential along the midplane. Predictions of the model are solid curves numerically calculated with Eqs.~(\ref{eq:p_ratio1}, \ref{eq:n_ratio}, \ref{eq:slopes}), and (\ref{eq:rrate2}). Markers are direct measurements of the respective quantities from simulations shortly after the reconnection rate peaks.}
    \label{fig:result}
\end{figure*}

Eq.~(\ref{eq:P2}) makes clear the factors affecting the x-line thermal pressure. $\langle \gamma(v_z)\rangle_{3}$ is controlled by the energy per particle that includes thermal motions in $v_z$. $\Gamma_y$ is purely controlled by the current carrier bulk flow. If all available energy is used to drive the current, then $\Gamma_y\simeq\langle \gamma(v_z)\rangle_{3}$, and $\left.P_{zz}\right|_{xline}$ becomes very small. Conversely, if only a small fraction of the total energy is needed to drive the current, $\left.P_{zz}\right|_{xline}$ can become significant. Approximating $\langle \gamma(v_z)\rangle_{3}\approx \langle \gamma(v_z)\rangle_{2}$ in the $\ell\rightarrow 0$ limit, we substitute Eq.~(\ref{eq:gamma}) for $\langle \gamma(v_z)\rangle_{3}$ and Eq.~(\ref{eq:gammay}) for $\Gamma_y$ to solve for the x-line pressure ratio:
\begin{equation}\label{eq:p_ratio1}  
    \frac{8\pi \left.P_{zz}\right|_{xline}}{B_{xm}^2} \leq2\frac{\left[2+(\sigma_m/2)\right](n_{3}^\prime/n_{1})-1}{\left[1+(\sigma_m/2)\right]\sqrt{1+(\sigma_m/2)(n_{1}/n_{3}^\prime)}}.
\end{equation}
The only free parameter in Eq.~(\ref{eq:p_ratio1}) is the proper compression ratio $n_{3}^\prime/n_{1}$. To determine this ratio, we first observe in simulations that the proper density around the x-line matches the inflow asymptotic density: $n_{3}^\prime\approx n_0$. Second, the upstream magnetic flux tube expands close to the diffusion region as the exhaust opens. The cold upstream plasma does not quickly redistribute itself along the expanding field lines. Thus local conservation of mass content within flux tubes implies that $n_0/n_{1}\approx B_{x0}/B_{xm}$ \citep{Li2021, Birn2009, Montag2017}. These two features are justified in Fig.~\ref{fig:ohms}(c). It follows that
\begin{equation}\label{eq:n_ratio}
    n_{3}^\prime/n_{1}\approx B_{x0}/B_{xm}.
\end{equation}
In the cold upstream plasma limit where $\left.P_{zz}\right|_{xline}\approx \left.P_{zz}\right|_{\delta}^0$, we close Eq.~(\ref{eq:p_ratio1}) using the following relations which were derived in \citet{Liu2022,Liu2017} based on the force balance along the inflow symmetry line and the assumption that separatrix field lines have the same slope, $S_{lope}$ (see Fig.~\ref{fig:cartoon}(a)), inside and outside the diffusion region:
\begin{subequations}\label{eq:slopes}
\begin{equation}\label{eq:slope1}
     S_{lope}^2 \approx 1-\frac{8\pi \left.P_{zz}\right|_{xline}}{B_{xm}^2}
\end{equation}
\begin{equation}
    \;\frac{B_{xm}}{B_{x0}}\approx \frac{1-S_{lope}^2}{1+S_{lope}^2}.
\end{equation}
\end{subequations}

{\it Results--}  
Using Eqs.~(\ref{eq:p_ratio1},~\ref{eq:n_ratio},~\ref{eq:slopes}) we numerically solve for the microscale and asymptotic x-line pressure ratios as a function of $\sigma_m$ or $\sigma_0$. Note that the relation between $\sigma_m$ and $\sigma_0$ is obtained numerically from the same equations.
The results are shown in Fig.~\ref{fig:result}, where we compare the theoretical predictions (solid curves) to a series of four 2.5D pair plasma PIC simulations (markers) with different initial magnetizations $\sigma_{0,\text{init}}$. Fig.~\ref{fig:result}(a) and (b) show excellent agreement between the simulation x-line pressure ratios and the predicted scaling relation with both $\sigma_m$ and $\sigma_0$.
In Fig.\ref{fig:result}(b), we find a simple scaling of $8\pi P_{zz}|_{xline}/B_{x0}^2\sim 2\sqrt{2/\sigma_0}$ in the $\sigma_0\gg 1$ limit (dashed curve). Note that fast reconnection is realized as long as $8\pi P_{zz}|_{xline}/B^2_{x0}<1$ \citep{Liu2020}.

We can also predict the reconnection rate from the x-line pressure ratio. The reconnection rate normalized to the asymptotic magnetic field is defined as $R_0\equiv cE_y/B_{x0}V_{A0}$, where $V_{A0}$ is the asymptotic upstream Alfv\'en speed. This can be written as 
\begin{equation}\label{eq:rrate}
    R_0 \simeq \left(\frac{B_{zm}}{B_{xm}}\right)\left(\frac{B_{xm}}{B_{x0}}\right)\left(\frac{V_{out,m}}{V_{A0}}
    \right)
\end{equation}
where $V_{out,m}$ is the outflow speed at the edge of the diffusion region. When $\sigma_m \gg 1$, $V_{out,m}\sim V_{A0}\sim c$ \citep{Liu2015}, and using $B_{zm}/B_{xm}\approx S_{lope}$ we can solve for the reconnection rate:
\begin{equation}\label{eq:rrate2}
    R_0\simeq S_{lope}\left(\frac{1-S_{lope}^2}{1+S_{lope}^2}\right)
\end{equation}
where $S_{lope}$ is calculated using Eqs.~(\ref{eq:slope1}) and (\ref{eq:p_ratio1},~\ref{eq:n_ratio}).
From Fig.~\ref{fig:result}(c), the model predicts a slower reconnection rate $R_0\sim\mathcal{O}(0.01)$ for $\sigma_0\sim\mathcal{O}(1)$, which rapidly increases to $R_0\sim0.1-0.3$ for $\sigma_0\gtrsim 10$. The prediction yields excellent agreement with our simulations and previously observed scaling \citep{Sironi2014}.

{\it Conclusions--} We have shown that in relativistic pair plasma reconnection, the x-line plasma pressure cannot balance the upstream magnetic pressure given the need to maintain the extreme x-line current density.   
In the absence of the Hall effect, essentially all inflowing electromagnetic energy near the x-line can be locally converted to plasma energy. However, current carriers constantly turn into the outflow. The energetic requirement to replenish these current carriers becomes significant at high $\sigma$ since the Lorentz factor associated with the current $\Gamma_y$ becomes extremely large. Therefore, only a small fraction of electromagnetic energy is available to maintain the x-line pressure $\left.P_{zz}\right|_{xline}$, making it significantly lower than the asymptotic upstream magnetic pressure $B_{x0}^2/8\pi$, leading to fast reconnection. We derived a simple scaling relation $8\pi P_{zz}|_{xline}/B_{x0}^2\sim 2\sqrt{2/\sigma_0}$ in the $\sigma_0\gg 1$ limit. The predicted reconnection rates agree well with PIC simulations.

While the reconnection outflow geometry opens out due to x-line pressure depletion in both astrophysical plasmas and non-relativistic electron-ion plasmas, the exhaust dynamics can be different. In the latter, the exhaust pressure can build up because Hall electromagnetic fields divert energy to the outflow region, and the energy required for the primary current carrier (electrons) is negligible \citep{Liu2022}. This results in a single, stable x-line. 
In contrast, pair plasma reconnection lacks the diversion of energy flow by Hall fields, and the relativistic current carriers take significant energy, causing low pressure within exhausts as well. Thus, the current layer can collapse even in the ``once opened'' exhaust region, explaining the bursty nature that recursively triggers and ejects magnetic islands \citep{Liu2020}. This time dependency, however, is not the driver and does not affect the average reconnection rate in collisionless plasmas.  
In comparison, the current in relativistic MHD models requires no energy, and the plasmoid instability was invoked to explain the fast rate \citep{Ripperda2019,Yang2019}. 

More work is needed to understand how this model couples with theories of turbulent driving and onset. While MHD-scale turbulence may enable fast reconnection to proceed independently of kinetic physics in the current sheet \citep{Lazarian2020}, some evidence shows that kinetic ($d_e$-scale) reconnecting layers persist and dominate a current sheet that is filled with external and self-generated turbulence \citep{Guo_2021}. Moreover, kinetic effects may enhance reconnection rates in the presence of broadband turbulence \citep{Haggerty2017}. In these cases, turbulent driving may lead to fast reconnection on kinetic scales, as detailed in this work. A resolution to this complex interplay requires separate dedicated efforts.

In summary, our model provides the critical theoretical foundation for fast reconnection in collisionless astrophysical plasmas. We expect these fundamental considerations of the current-carrier requirement and x-line energy budget to carry over to three-dimensional (3D) systems. \\



\begin{acknowledgments}
MG thanks the Wilder Fellowship at Dartmouth College and NSF Grant PHY- 1902867 for support. YL is grateful for supports from NSF grant PHY- 1902867 through the NSF/DOE Partnership in Basic Plasma Science and Engineering, NSF Career Award 2142430 and NASA's MMS mission 80NSSC21K2048. We thank Xiaocan Li for helpful discussions and assistance with simulations. Simulations were performed at the National Energy Research Scientific Computing Center at LBNL.
\end{acknowledgments}

\appendix 

{\it Appendix on the derivation of the relativistic Ohm's law and the diffusion region width--}
The following derivation of the relativistic Ohm's law is due to \citet{Zenitani2018}. The stress-energy tensor can be decomposed as 
\begin{equation}\label{se_decomp}
    T^{\alpha\beta}=\mathcal{E}U^\alpha U^\beta+Q^{\alpha\beta}+ P^{\alpha\beta}
\end{equation}
where $U^\alpha$ is an arbitrary flow 4-velocity. With $\Delta^{\alpha\beta}\equiv \eta^{\alpha\beta}+U^\alpha U^\beta$ as the projection tensor, $\mathcal{E}\equiv T^{\alpha\beta}U_\alpha U_\beta$ is the energy density in the $U^\alpha$-moving frame. $Q^{\alpha\beta}=q^\alpha U^\beta + U^\alpha q^\beta$ is the heat flux tensor, where $q^\alpha \equiv -\Delta_\beta^\alpha T^{\beta\mu}U_\mu$ is the heat flux 4-vector. $P^{\alpha\beta}\equiv\Delta^\alpha_\mu\Delta^\beta_\nu T^{\mu\nu}$ is the pressure tensor projected in the $U^\alpha$-moving frame.

With $U^\alpha = U_E^\alpha$ (the Eckart 4-velocity), the energy-momentum equation for the electron species is then given by 
\begin{equation}\label{eq:electron_EM}
    \partial_\beta T^{\alpha\beta}_{(-)} = -e n^\prime F^{\alpha\beta}U_\beta
\end{equation}
where $F^{\alpha\beta}$ is the electromagnetic tensor.
The relativistic electron Ohm's law is obtained from the space components of Eq.~(\ref{eq:electron_EM}):
\begin{equation}
    \mathbf{E}+\mathbf{V}\times\mathbf{B}=-\frac{1}{\Gamma e n^\prime}\left[\partial_j\left(\mathcal{E}U^i U^j+Q^{ij}+P^{ij} \right)+ \partial_t T^{i0} \right].
\end{equation}
Here $\mathbf{V}$ is the electron Eckart 3-velocity, and $\Gamma$ is the Lorentz factor of the Eckart frame. 

As discussed the main text, it is known that in the relativistic regime, the bulk inertia term of the Ohm's law balances the reconnection electric field $E_y$ at the edge of the diffusion region \citep{Hesse2007,Zenitani2018,Xiong}. Note that the edge of the diffusion region coincides with the edge of the current sheet.
Thus we evaluate Ohm's law at the transition region close to surface 1 as follows: 

\begin{equation}\label{eq:Ohm1}
    E_{y}=-V_{z1}B_{xm}\approx-\left. \frac{1}{\Gamma e n^\prime}\partial_j\left(\mathcal{E} U_y U^j\right)\right|_1.
\end{equation}
At the very edge of the current sheet, the plasma has not yet been accelerated and is essentially in its upstream state, where $\Gamma_1 \approx 1$, $U_{y1}\approx 0$, and the internal energy $\mathcal{E}_1\approx n'_{1}mc^2$. In addition, the z-direction derivative dominates, so that $\partial_j\left(\mathcal{E} U_y U^j\right)\approx \partial_z\left(\mathcal{E} U_y U_z\right)$. Since $U_{y1}\approx 0$, the product rule  implies that 
\begin{equation}
    \left.\partial_z\left(\mathcal{E}U_yU_z\right)\right|_1\\=\left.\mathcal{E}U_z(\partial_zU_y) \right|_1=\left.\mathcal{E}\Gamma V_{z}(\partial_zU_y) \right|_1
\end{equation}
With $U_{z1}=\Gamma_1 V_{z1}$, we can rewrite Eq.~(\ref{eq:Ohm1}) as
\begin{equation}
    B_{xm}\approx \frac{mc^2}{e}\left.\left(\partial_z U_y\right)\right|_1.
\end{equation}
At the transition region, $U_y$ is very small but has begun to increase towards its peak value at the center of the current sheet.
Hence $\left(\partial_z U_y\right)|_1\approx -U_{y3}/\delta = J_{y3}/2en^\prime_{3}\delta$. 

Finally, use Ampere's law $B_{xm}/4\pi\delta=J_{y3}$ to conclude that the current sheet width is given by the pair plasma inertial length evaluated inside the current sheet:
\begin{equation}
    \delta \approx \sqrt{\frac{mc^2}{8\pi n_{3}^\prime e^2}}.
\end{equation}

{\it Appendix on the simulation details--} Simulations are performed with VPIC, which evolves particles with the relativistic Vlasov equation and fields with Maxwell's equations \citep{Bowers2008}. We test initial asymptotic magnetizations of $\sigma_{0,\text{init}}=10$, $100$ , $500$, and $1000$. All runs have $x\times z$ system size $1086\,d_e\times1086\,d_e$ and grid size $6144\times 6144$, except for the $\sigma_{0,\text{init}}=1000$ run which has grid size $6144\times 12288$. The initial configuration is a symmetric force-free current sheet without background guide field. The field profile is $\mathbf{B}_0=B_{x0}\left[\mathrm{tanh}(z/\lambda)\hat{\mathbf{x}} + \mathrm{sech}(z/\lambda)\hat{\mathbf{y}}\right]$. The initial current sheet half-width $\lambda=0.85\sqrt{\sigma_{0,\text{init}}}d_e$, where $d_e\equiv(mc^2/8\pi n_0 e^2)^{1/2}$ is the upstream pair plasma inertial length.  Reconnection is initiated with a small magnetic perturbation. Both species have initial temperature $T_0=0.5 mc^2$, and there are around $7.5$ billion macroparticles.

\providecommand{\noopsort}[1]{}\providecommand{\singleletter}[1]{#1}%

\end{document}